\documentclass[runningheads]{llncs}
\usepackage{graphicx}
\usepackage{hyperref}

\usepackage{ctable} 
\usepackage{fancyvrb}

\usepackage{multirow} 
\usepackage{float}
\usepackage{threeparttable}
\usepackage{pifont}
\usepackage{array}
\newcommand{\PreserveBackslash}[1]{\let\temp=\\#1\let\\=\temp}
\newcolumntype{C}[1]{>{\PreserveBackslash\centering}p{#1}}
\newcolumntype{R}[1]{>{\PreserveBackslash\raggedleft}p{#1}}
\newcolumntype{L}[1]{>{\PreserveBackslash\raggedright}p{#1}}
\usepackage{listings}
\usepackage{algorithm}
\usepackage[noend]{algpseudocode}

\makeatletter
\renewcommand{\ALG@name}{Code}
\makeatother
\lstdefinelanguage{Fortran}
{
	keywords = {parallel, loop, if, in, while, do, else,  enddo, end, concurrent},
	comment = [l]{//},
}
\begin{document}
\title{Can Fortran's `{\tt do concurrent}' replace directives for accelerated computing?\thanks{Supported by NSF awards AGS 202815 and ICER 1854790, and NASA grant 80NSSC20K1582. This work used the Extreme Science and Engineering Discovery Environment (XSEDE) Bridges2 at the Pittsburgh Supercomputer Center through allocation TG-MCA03S014.  It also used the DGX A100 system at the Computational Science Research Center at San Diego State University provided by NSF award OAC 2019194}}
\titlerunning{{Can \tt {\tt do concurrent}} replace directives?}
\author{Miko M. Stulajter\orcidID{0000-0003-0939-1055}, Ronald M. Caplan\orcidID{0000-0002-2633-4290}   \and Jon A. Linker\orcidID{0000-0003-1662-3328} }
\authorrunning{M. Stulajter, R. M. Caplan, et al.}
%
\institute{Predictive Science Inc. 9990 Mesa Rim Road Suite 170, San Diego, CA  92121 \email{\{miko,caplanr,linkerj\}@predsci.com}\\
\url{http://www.predsci.com}}
\maketitle
\begin{abstract}
Recently, there has been growing interest in using standard language constructs (e.g. C++'s Parallel Algorithms and Fortran's {\tt do concurrent}) for accelerated computing as an alternative to directive-based APIs (e.g. OpenMP and OpenACC). These constructs have the potential to be more portable, and some compilers already (or have plans to) support such standards.
Here, we look at the current capabilities, portability, and performance of replacing directives with Fortran's {\tt do concurrent} using a mini-app that currently implements OpenACC for GPU-acceleration and OpenMP for multi-core CPU parallelism.  We replace as many directives as possible with {\tt do concurrent}, testing various configurations and compiler options within three major compilers: GNU's {\tt gfortran}, NVIDIA's {\tt nvfortran}, and Intel's {\tt ifort}.
We find that with the right compiler versions and flags, many directives can be replaced without loss of performance or portability, and, in the case of {\tt nvfortran}, they can all be replaced.  We discuss limitations that may apply to more complicated codes and future language additions that may mitigate them.  
The software and Singularity containers are publicly provided to allow the results to be reproduced.
\keywords{accelerated computing \and OpenMP \and OpenACC \and do concurrent \and standard language parallelism.}
\end{abstract}

\section{Introduction}
\label{sec:intro}
OpenMP\footnote{\url{www.openmp.org}}\cite{van2017using} and OpenACC\footnote{\url{www.openacc.org}}\cite{chandrasekaran2017openacc} are popular directive-based APIs for parallelizing code to run on multi-core CPUs and GPUs.  For accelerated computing, they provide a higher-level approach to accelerating codes without requiring writing specialized low-level, often vendor-specific, API code (e.g. CUDA, ROCm, OpenCL, etc.).   Since they mostly consist of specialized comments/pragmas, they exhibit backward compatibility, allowing a non-supported compiler to simply ignore them and still compile the code as before.   This makes directive-based approaches very desirable for legacy codes, and helps to allow compartmentalized development.  However, they also can suffer from incomplete vendor, hardware, and/or compiler support, make codes somewhat harder to read, and, due to their rapid development, are less future-proof than standard languages, possibly requiring occasional re-writes.


Due to the widespread adoption of multi-core CPUs and accelerators, standard languages have begun to add built-in features that may help/enable compilers to parallelize code.   This includes C++17's Standard Parallel Algorithms and Fortran's {\tt do concurrent} (DC) (see Refs.~\cite{stdparcpp,stdpardc} for examples using the NVIDIA HPC SDK\footnote{\url{https://developer.nvidia.com/hpc-sdk}}).  Standard parallel language features have the potential to remove the need for directives, making multi-threaded and accelerated codes fully portable across compiler vendors and hardware.  However, this requires compiler support, and few have been quick to implement these features for GPU acceleration.  

Here, we focus on Fortran's DC construct.  The NVIDIA HPC SDK is the only compiler at the time of this writing with accelerator support using DC, while Intel has indicated plans to add such support in an upcoming release of their {\tt ifort} compiler included in the OneAPI HPC Toolkit\cite{inteldc}.  Other compilers that support directive-based accelerator offloading in Fortran include GCC's {\tt gfortran}\footnote{\url{https://gcc.gnu.org/}}, LLVM {\tt flang}\footnote{\url{https://flang.llvm.org}}, AOCC's extended {\tt flang}\footnote{\url{https://developer.amd.com/amd-aocc}}, IBM's {\tt XL}\footnote{\url{https://www.ibm.com/products/xl-fortran-linux-compiler-power}}, and HPE's Cray Fortran\footnote{\url{https://support.hpe.com/hpesc/public/docDisplay?docId=a00115296en_us&page=index.html}}, but we could not find any announced plans for these to support DC for accelerated computing in the near future.

In this paper, we investigate the current capabilities, portability, and performance of replacing directives with DC in a Fortran mini-app that currently implements directives for GPU-acceleration and multi-core CPU parallelism.  We replace as many directives as possible with DC, testing various run-time configurations and compilers.  Our mini-app currently uses OpenACC with either {\tt nvfortran} or {\tt gfortran} for GPU-acceleration on NVIDIA GPUs, and uses OpenMP with {\tt nvfortran}. {\tt gfortran}, or {\tt ifort} for multi-core CPU parallelism (as well as OpenACC multi-core with {\tt nvfortran}).  A key portability concern is if replacing directives with DC for GPU-acceleration will result in a loss of multi-core CPU parallelism.  Therefore, we test if each compiler can parallelize the DC loops for multi-core CPUs.  We note that for codes using non-hybrid MPI for CPU parallelism and MPI+OpenMP/ACC for GPU acceleration, this is not as much of a concern.

The paper is organized as follows: In Sec.~\ref{sec:mini-app}, we describe our Fortran mini-app with its current directive-based parallelelization, along with the test case we use, showing baseline performance results.  In Sec.~\ref{sec:implementation}, we describe the implementation of DC into the mini-app, first introducing its capabilities and support, and then showing examples of replacing OpenMP/ACC directives with DC, including a discussion of current limitations.  Then the resulting mini-app source code versions and compiler flag options used for the tests are described.  Performance and compatibility results are reported in Sec.~\ref{sec:results} for both multi-core CPU and GPU runs.  Finally, discussion on the current status of DC and its potential to replace directives is given in Sec.~\ref{sec:discussion}.  Instructions on how to access and use our provided Singularity containers and codes to reproduce the results in the paper are given in the Appendix.
	

\section{Code and test description}
\label{sec:mini-app}
To investigate the current capabilities, portability, and performance of replacing directives with DC, we use a Fortran mini-app called {\tt diffuse} that currently implements directives for GPU-acceleration and multi-core CPU parallelism.  Here we describe the code, the test case we use, the computational test environment, and baseline performance results.

\subsection{Code description}
NASA and NSF have recently supported a program called "Next Generation Software for Data-driven Models of Space Weather with Quantified Uncertainties", whose main objective is to improve predictions of solar wind and coronal mass ejections to investigate how they might impact Earth.  This will be done by developing a new data-driven time-dependent model of the Sun's upper atmosphere.  One key component of this model is the use of a data-assimilation flux transport model to generate an ensemble of magnetic field maps of the solar surface to use as boundary conditions.   To accomplish this, we have been developing  an Open-source Flux Transport (OFT) software suite, whose key computational core is the High-Performance Flux Transport code (HipFT).  HipFT currently implements OpenACC for GPU-acceleration and OpenMP for multi-core CPU parallelism, and we are interested in replacing the directives with DC.

In order to test the use of DC, we use a mini-app called {\tt diffuse} that implements the most computationally expensive algorithm (surface diffusion) of the flux transport in HipFT. {\tt diffuse}'s source code for the diffusion algorithm is identical to that of HipFT.  The diffusion algorithm integrates a spherical surface heat equation on a logically rectangular non-uniform grid.   The operator is discretized with a second-order central finite-difference scheme in space, while the time integration uses the second-order Legendre polynomial extended stability Runge-Kutta scheme (RKL2) \cite{RKL2_2014,Caplan_2017}.  

\subsection{Test description}
\label{sec:testcase}
\begin{figure}
    \centering
    \includegraphics[width=0.49\textwidth]{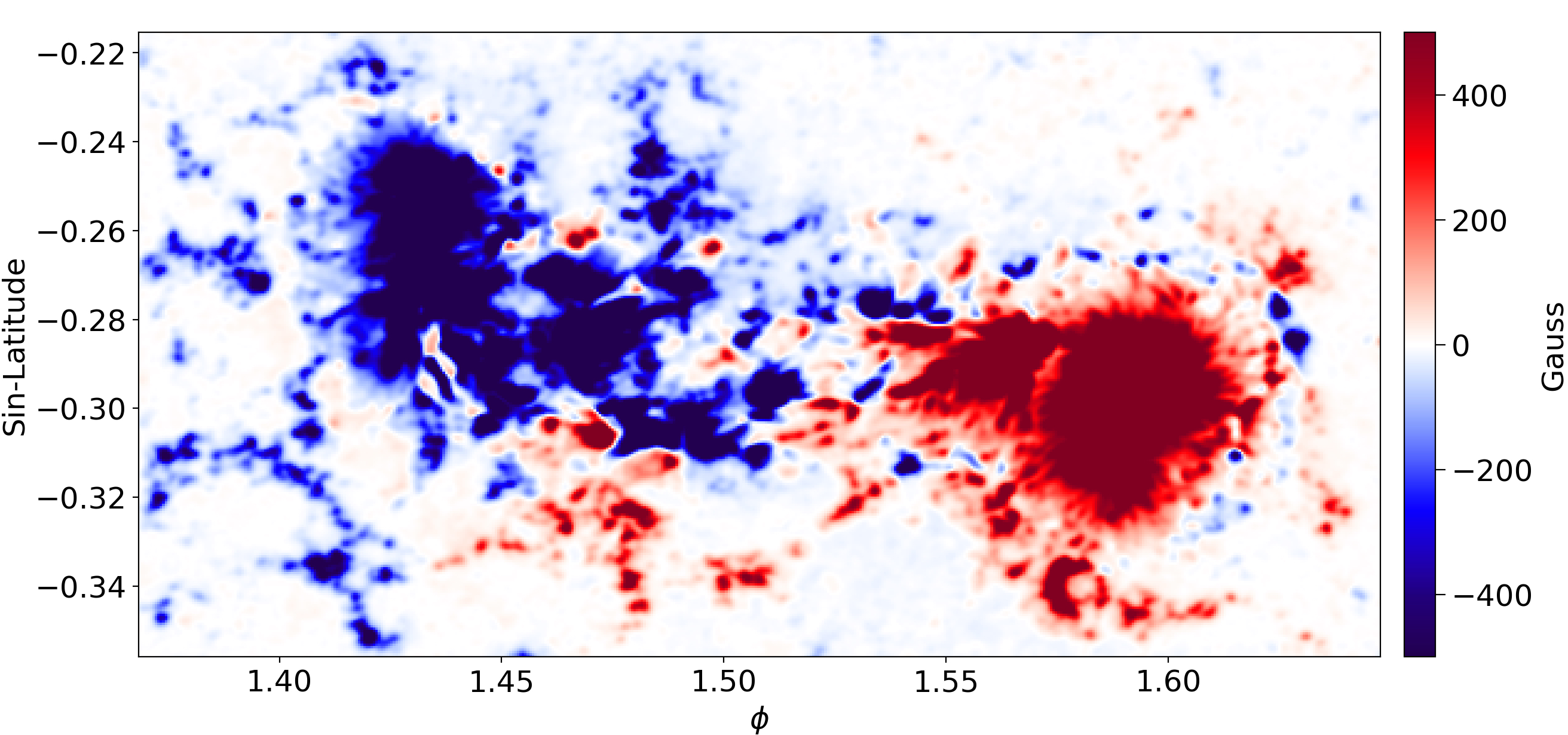}
    \includegraphics[width=0.49\textwidth]{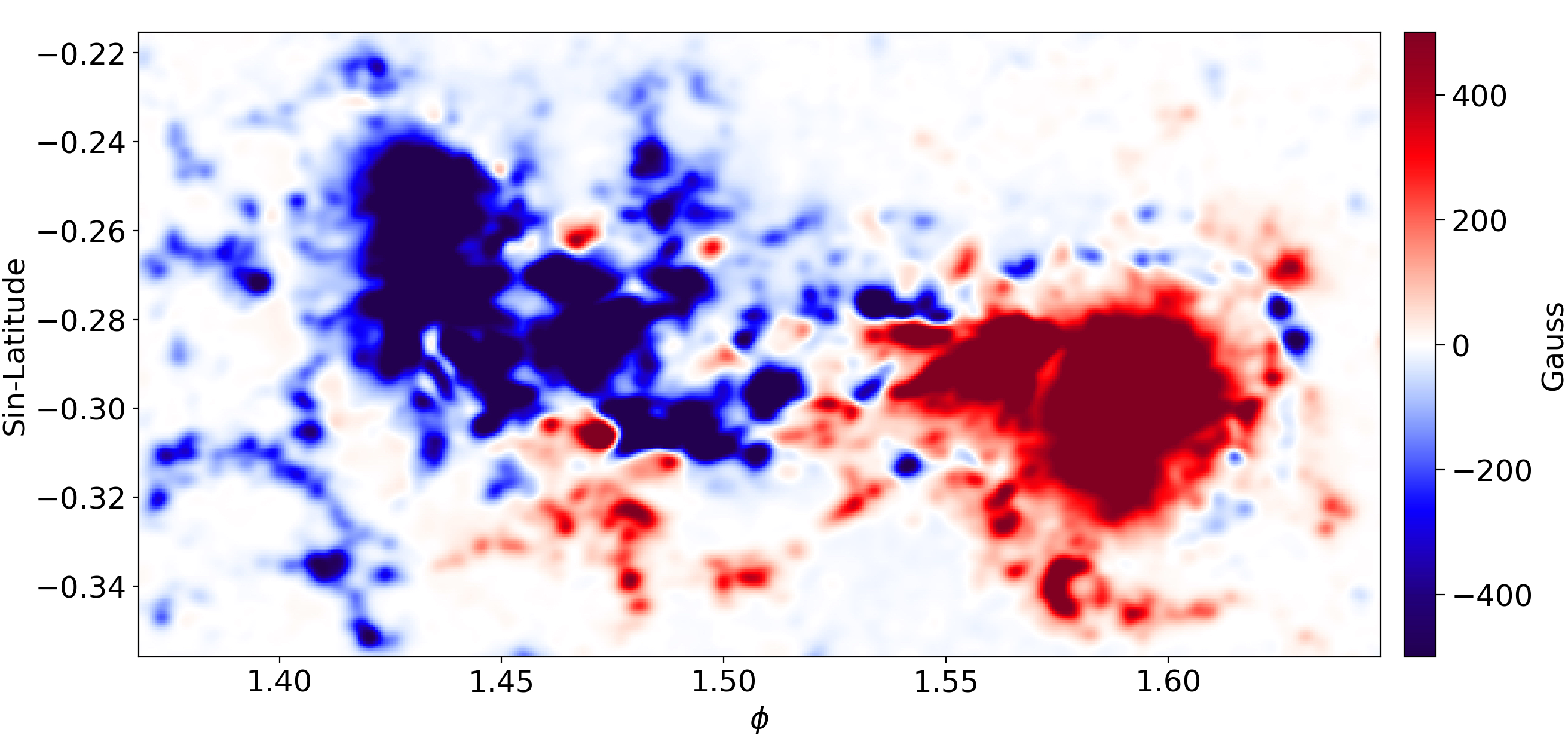}
    \caption{Zoomed-in detail of the test case magnetic field map before (left) and after (right) smoothing with {\tt diffuse}.}
    \label{fig:testcase}
\end{figure}

Although {\tt diffuse} is used here as a mini-app representation of HipFT, it is also used in production to slightly smooth solar surface magnetic fields to prepare them for use in models of the corona \cite{mapprep}.  As such, we select a real-world example of using {\tt diffuse}, that of smoothing the `Native res PSI map` described in Ref.~\cite{Caplan_2021}.  This large map has a resolution of $3974\times2013$ in $(\theta,\phi)$ and takes 40,260 total iterations of applying the diffusion operator to smooth.  A detail from the map before and after running {\tt diffuse} is shown in Fig.~\ref{fig:testcase}.

\subsection{Computational environment}
In order to best assess the capabilities of the compiler support for DC, we use the latest available versions of the compilers at the time of testing.  These are shown in Table~\ref{tab:compilers}.
\begin{table}[htbp]
\begin{center}
\caption{Compiler versions used in tests. \label{tab:compilers}}
\begin{tabular}{||L{45mm}  |C{25mm} |C{15mm}||}
\hline
\multicolumn{1}{||c|}{Compiler Suite} & Compiler & Version \\
\hline
\hspace*{4pt} GNU Compiler Collection & {\tt gfortran} & 11.2 \\
\hspace*{4pt} NVIDIA HPC SDK &  {\tt nvfortran} & 21.7 \\
\hspace*{4pt} Intel OneAPI HPC Toolkit & {\tt ifort} (classic) & 21.3 \\
\hline
\end{tabular}
\end{center}
\end{table}
The CPU tests are run on the Bridges2 system located at the Pittsburgh Supercomputing Center using our allocation obtained through NSF's XSEDE program\cite{xsede}.  The GPU tests are run on an NVIDIA DGX A100 server at San Diego State University.  Since {\tt diffuse} does not have multi-node or multi-GPU capabilities, the CPU tests are run on a single CPU node, while the GPU tests are run on a single GPU within the DGX system.  The hardware specifications are shown in Table~\ref{tab:hwspecs}.
\begin{table}[htbp]
\begin{center}
\caption{Hardware utilized for all test runs. \label{tab:hwspecs}}
\begin{tabular}{||L{44mm}  |C{38mm} |C{34mm}||}
\hline
\, & CPU & GPU \\
\hline
\multirow{2}{*}{\hspace*{4pt}CPU/GPU Model} & (2x) AMD EPYC 7742 & NVIDIA A100\\
& (128 cores) & SXM4 \\
\hspace*{4pt}Peak Memory Bandwidth & 381.4 GB/s  & 1555 GB/s         \\
\hspace*{4pt}Clock Frequency (base/boost)  & 2.3/3.4 GHz  & 1.1/1.4 GHz         \\
\hspace*{4pt}RAM             & 256 GB     & 40 GB            \\
\hspace*{4pt}Peak DP FLOPs   & 7.0 TFLOPs & 9.8 TFLOPs       \\
\hline
\end{tabular}
\end{center}
\end{table}

Since systems do not always have the latest compilers available, and setting up our code's dependencies can be difficult, we utilize Singularity containers\cite{kurtzer_sochat_bauer}.  These containers are built with the compiler environment and our dependent libraries pre-installed so they can be easily used to build and run the code.  We use Singularity 3.8.0, and for GPU runs, use the `-{}-nv' flag to connect to the NVIDIA driver (and CUDA library) on the local system.  The CUDA run-time library used for GPU runs was version 11.4.  As shown in the Appendix, running the codes in the containers yields virtually the same performance as a bare metal installation.  All the test runs performed in this paper can be reproduced using the containers along with the code, both of which are publicly released in Ref.~\cite{mikic_zoran_2021_5253520} and at \url{www.predsci.com/papers/dc}.

\subsection{Baseline performance results}
\label{sec:baseline}
\begin{figure}
    \centering
    \includegraphics[height=1.75in]{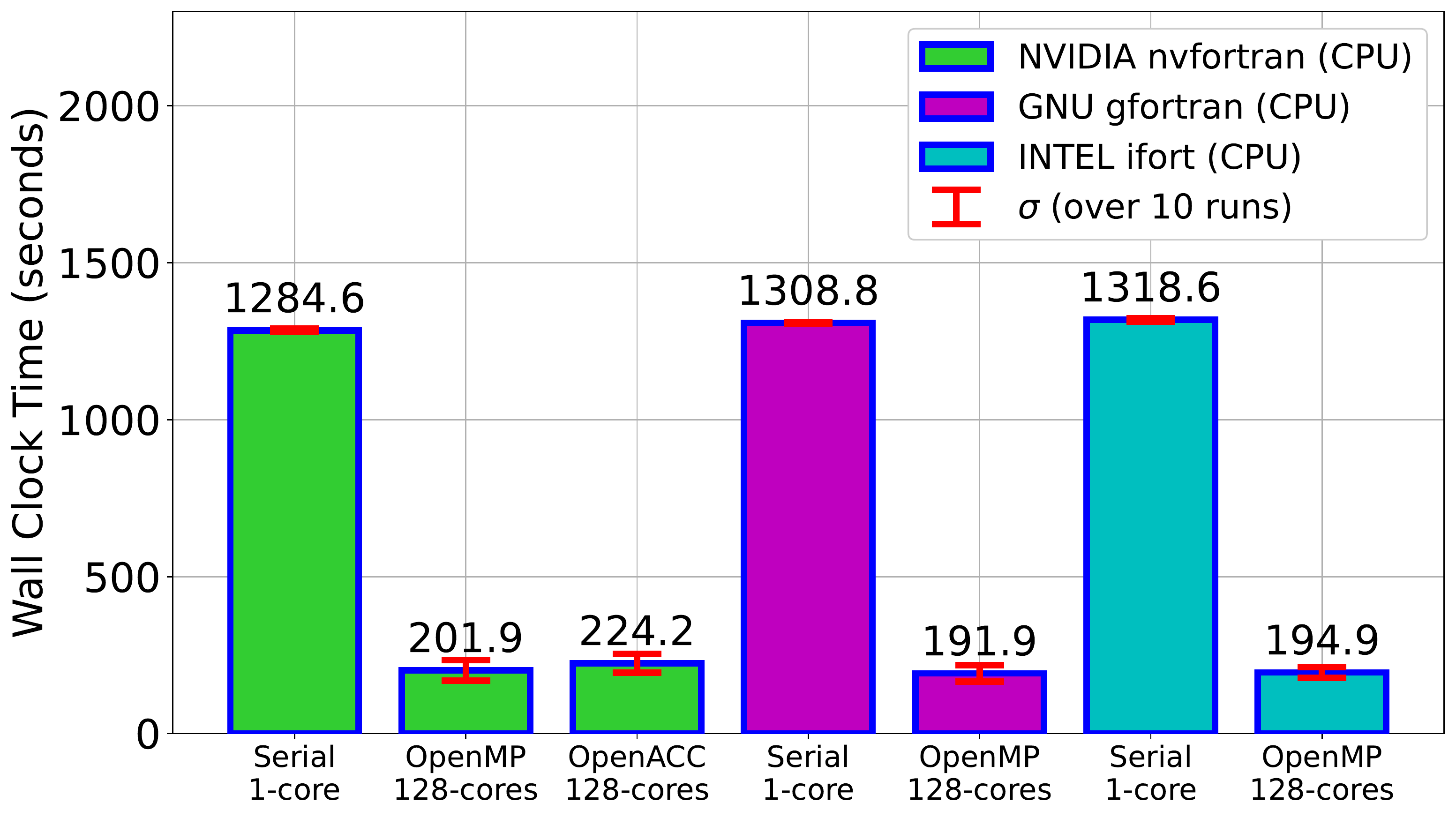}
    \includegraphics[height=1.775in]{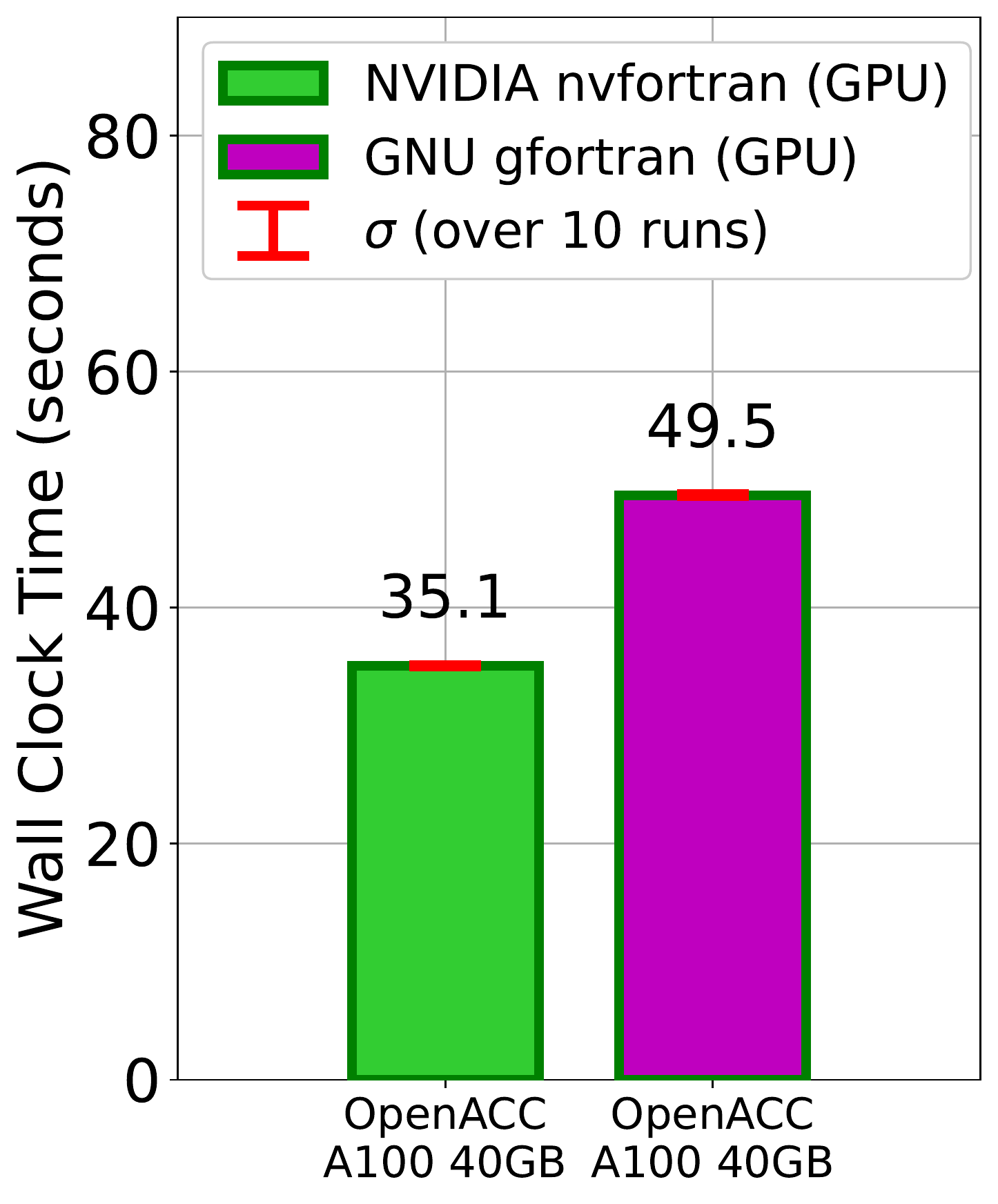}
    \caption{Baseline CPU and GPU timing results of the original {\tt diffuse} code run on the test case.  Times shown are averages over 10 runs, and the standard deviations are shown.}
    \label{fig:baseline}
\end{figure}

Our goal in this paper is to test replacing directives with DC for accelerated computing, ensuring we do not lose multi-core CPU parallelism, and that the performance is comparable to the original directive-based code. It is not our focus to compare performance between  the various compilers and hardware.  We therefore use similar basic compiler optimization flags (shown in Sec.\ref{sec:codelist}) for each compiler-hardware combination and do not explore all possible optimizations.  In order to compare the performance of the original code to the modified versions, we perform baseline timings of the original code.  For these, and all timing results in the paper, we run each test 10 times and take the average of the full wall clock times (which include all I/O and GPU-CPU data transfer time). In Fig.~\ref{fig:baseline} we show the baseline timings along with their standard deviations.   We also include CPU runs on a single CPU core (serial) to illustrate the multi-core CPU parallelism.
We see that each compiler obtains comparable performance on the CPU runs, yielding a speedup of $\sim 7\times$ when using 128 CPU cores compared to running in serial.  While this may seem low, it is common for highly memory-bound algorithms to exhibit such non-ideal single node multi-threaded scaling \cite{balarac2020avbp}.  The performance of the {\tt nvfortran} CPU run using OpenMP is $\sim 10\%$ faster than using OpenACC for multi-core parallelism.  On the GPU, the {\tt nvfortran} OpenACC GPU run is $\sim 30\%$ faster than the {\tt gfortran} run, which is not unexpected considering {\tt nvfortran} has a more mature implementation of OpenACC.


\section{Implementation}
\label{sec:implementation}
In this section, we first give a background on Fortran's {\tt do concurrent} construct, and then describe our implementation of it into the {\tt diffuse} code, and the resulting code variations.  We also describe the compiler flags used for each code version and compiler combination.

\subsection{The Fortran {\tt do concurrent} construct}
In 2008, ISO Standard Fortran introduced the DC construct for loops as an alternative to the standard {\tt do} loop (or nested  {\tt do} loops).  DC indicates to the compiler that the loop's iterations can be computed in any order.  This potentially allows for the expression of parallelism of loops directly in the Fortran language, making it easier for compilers to parallelize the loops.   While any-order execution is a necessary condition for parallelization, it is not always sufficient (for example,  reduction and atomic operations, as well as others\footnote{\url{https://releases.llvm.org/12.0.0/tools/flang/docs/DoConcurrent.html}}).  Therefore, DC can be viewed as providing a hint to the compiler that the loop is likely able to be parallelized.  Work in helping make DC fully sufficient for parallelism is on-going, with Fortran 2018 adding locality statements (allowing specification of private and shared variables \footnote{\url{https://j3-fortran.org/doc/year/18/18-007r1.pdf}}), and specifying reductions in DC will be included in the upcoming Fortran 202X release\footnote{\url{https://j3-fortran.org/doc/year/21/21-007.pdf}}.

The syntax of non-nested {\tt do} loops and {\tt do concurrent} loops are similar. A {\tt do} loop has the syntax {\tt do index=start,end} while a {\tt do concurrent} loop has the syntax {\tt do concurrent (index=start:end)}. The only key difference is the addition of the word {\tt concurrent} and a small change to the loop parameters where there is the addition of parentheses and a replacement of the comma with ellipses. With nested loops, there is more of a difference in formatting. Code \ref{Alg_1} shows  nested {\tt do} loops parallelized with directives. The loop nest is shown with both OpenMP and OpenACC directives in the manner they are used in the mini-app. This nested {\tt do} loop example spans 8 lines with directives, but can be written in 3 lines with DC as shown in Code \ref{Alg_2}. With DC loops, nested loops initialization statements are collapsed into one initialization statement. The syntax of DC loops is as follows: {\tt do concurrent (index1=start1:end1, index2=start2:end2, ...)}. As this example shows, DC loops make nested {\tt do} loops more compact and easier to read. 
\begin{algorithm}
		\caption{ Nested {\tt do} loops with OpenMP/ACC directives}
		\label{Alg_1}
		\begin{algorithmic}
			\State !\$omp parallel do collapse(2) default(shared)
			\State !\$acc parallel loop collapse(2) default(present)
			\State \textbf{\hspace*{20pt} do} { i=1,N}
			\State \textbf{\hspace*{40pt} do} {  j=1,M}
			\State \hspace*{60pt} Computation
			\State \textbf{\hspace*{40pt} enddo} 
			\State \textbf{\hspace*{20pt} enddo} 
			\State !\$acc end parallel loop
			\State !\$omp end parallel do
		\end{algorithmic}
\end{algorithm}
\begin{algorithm}
		\caption{ Nested {\tt do} loops as a {\tt do concurrent} loop}
		\label{Alg_2}
		\begin{algorithmic}
			\State \textbf{\hspace*{20pt} do concurrent} { (i=1:N, j=1:M)}
			\State \hspace*{40pt} Computation
			\State \textbf{\hspace*{20pt} enddo} 
		\end{algorithmic}
\end{algorithm}

Most current compilers support the Fortran 2008 standard, which includes the basic DC syntax.  However, since the specification does not require that the compiler try to parallelize the loops, they are often treated as serial {\tt do} loops.  When a compiler does support parallelization of DC, special compiler flags are needed to activate the feature (see Sec.~\ref{sec:flaglist} for details).   

Although the latest version of the OpenACC (3.1)\footnote{\url{https://www.openacc.org/blog/announcing-openacc-31}} specification adds support for decorating DC loops with directives, at present, there are no implementations of this support (with the possible exception of using the {\tt kernels} directive).  There is also no mention of supporting directives on a DC construct within the most recent OpenMP (5.1)\footnote{\url{https://www.openmp.org/spec-html/5.1/openmp.html}} specification.  Therefore, replacing {\tt do} loops with DC may break the ability to parallelize the loops when using compilers that do not support direct DC parallelization.  

The current state of DC support is varied across different compilers and versions. {\tt nvfortran} 18.1 added serial support for DC along with locality of variables, while {\tt nvfortran} 20.11 added support for parallelization of DC loops for both CPUs and GPUs.   In {\tt gfortran} 8, serial support for DC was introduced, while {\tt gfortran} 9 added support for parallelization of DC on multi-core CPUs using {\tt gfortran}'s auto parallelization feature. {\tt ifort} started supporting serial DC loops in version 12.  Then, in version 16, parallelization support was added through {\tt ifort}'s auto parallelization feature (using the flag {\tt -parallel}). With version 19.1, locality of variable support was added, and parallelization became linked to the OpenMP compiler flags. Table \ref{support_DC} gives a summary of the current support of parallel DC loops for the compilers used in this paper.
\begin{table}[htbp]
	\vspace{-4mm}
	\caption{Current support of DC loop parallelization for the compilers used in this paper.} 
	\label{support_DC}
	\begin{center}
		\begin{tabular}{||C{18mm}| C{15mm} | C{82mm}||} 
				\hline
				Compiler & Version & {\tt do concurrent} parallelization support\\
				\specialrule{.1em}{.05em}{.05em}
				\multirow{2}{*}{{\tt gfortran}}
					&\multirow{2}{*}{$\ge9$} &  Parallelizable on CPU with {\tt -ftree-parallelize-loops=<N>}\\
					&& flag. Locality of variables is \emph{not} supported. \\
				\hline
				\multirow{2}{*}{{\tt nvfortran}}
			 		&\multirow{2}{*}{$\ge20.11$}&  Parallelizable on CPU and GPU with the {\tt -stdpar} \\
			 		&& flag. Locality of variables is supported. \\
			 	\hline
			 	\multirow{2}{*}{{\tt ifort}}
			 		& \multirow{2}{*}{$\ge19.1$} & Parallelizable on CPU with the {\tt -fopenmp} flag. \\
			 		&& Locality of variables is supported. \\
			\hline			
		\end{tabular}
	\end{center}
	\vspace{-8mm}
\end{table}

\subsection{Code Versions}
\label{sec:codelist}
Here we list the code variants that we use to test the portability and performance of replacing directives with DC in {\tt diffuse}.   For versions that use DC, only basic DC loop syntax was used with no locality of variables, as not all compilers support this feature in all configurations.

\emph{Original}: 
This is the original version of {\tt diffuse} which uses OpenACC and OpenMP directives on all parallelizable {\tt do} loops as well as OpenACC data movement directives.  It does not contain any DC loops.  It is the code version used for the performance results of Sec.~\ref{sec:baseline}, and will be the standard we compare to for both performance and compatibility.

\emph{New}: 
This version is obtained by replacing directive surrounded {\tt do} loops in \emph{Original} with DC loops, with the exception of reduction loops.  The directives on the reduction loops are kept since reductions are not supported in parallelized DC loops (see discussion in \emph{Experimental}).  We also keep all OpenACC data directives for explicit GPU data management.  This code is expected to perform as well as the \emph{Original} code if the DC loops are recognized and implemented efficiently. 

\emph{Serial}: 
This version contains no OpenACC or OpenMP directives at all, nor any DC loops.  It is the same as \emph{Original} with all directives removed.  It should run in serial in all cases, unless an auto-parallelizing feature of a compiler is utilized.  We include this code as a control and to ensure the multi-core CPU parallel runs are exhibiting the expected parallelism.

\emph{Experimental}:  
This version does not contain any OpenMP or OpenACC directives at all, replacing all loops (including reduction loops) with DC.  A key feature of this code version is that it represents the `ideal' scenario of using only the Fortran standard language for accelerated computing without needing any directives.  This version does not technically violate the Fortran standard since a DC on a reduction loop is valid if not parallelized, as the iterations can be computed in any order.  However, if the compiler does  parallelize these DC reduction loops, it will likely produce wrong results due to race conditions, unless it supports implicit analysis and implementation of DC reductions.   As mentioned above, Fortran 202X will add reductions to DC, resolving this problem.   Removing all directives also removes explicit GPU-CPU data movement, whose absence will lead to very poor performance on accelerators (due to repeated data movement between the CPU and GPU) unless the compiler supports automatic GPU-CPU memory management.  Features such as NVIDIA's Unified Memory and AMD's Smart Access Memory can allow compilers to resolve this issue.  

In Table \ref{Code_versions} we summarize all versions of the code we use for our tests.
\begin{table}[htbp]
	\vspace{-4mm}
	\caption{Summary of DC and directive implementations for each version of the {\tt diffuse} code tested.}
	\label{Code_versions}
	\begin{center}
		\begin{tabular}{||C{21mm}  |C{40mm} |C{54mm}  ||}
			\hline
			\, & {\tt do concurrent}  & Directives \\
			\hline
			\emph{Original} & None &  all loops \& data management \\
			\emph{New}  &   all loops except reductions &   reduction loops \& data management\\
			\emph{Serial} & None & None \\
			\emph{Experimental} & all loops & None \\
			\hline		
		\end{tabular}
	\end{center}
	\vspace{-8mm}
\end{table}
	
\subsection{Compiler Flag Options}
\label{sec:flaglist}
The {\tt gfortran}, {\tt nvfortran}, and {\tt ifort} compilers each have different flags to implement code parallelization and optimizations.  Here we describe the compiler flags we use for each code version, compiler, and target hardware configuration.  For all compilers, we use the {\tt -O3} flag to activate typical compiler optimizations, and {\tt -march=<ARCH>} to tell the compiler to target the specific CPU we run the tests on.  Typically, we use {\tt native} for {\tt <ARCH>} to automatically target the current system, but some configurations (such as using {\tt ifort} on AMD EPYC CPUs) required us to specify the option manually (in that case {\tt <ARCH>} is set to {\tt core-avx2}).  All \emph{Serial} code versions use only these default compiler flags.

\emph{\tt nvfortran}:   
For GPU parallelization, the \emph{Original} code uses the flag {\tt -acc=gpu} which enables the OpenACC directives.   We also include the flag {\tt -gpu=ccXY, cudaX.Y} to specify the specific GPU run time and hardware capabilities (similar to {\tt -march} for CPUs). The {\tt ccXY} indicates a device with compute capabilities of {\tt X.Y}, while {\tt cudaX.Y} tells the compiler to use the {\tt X.Y} version of the CUDA library.  To check if/how the compiler parallelized the loops, we set {\tt -Minfo=accel}, which outputs parallelization information.  

For the \emph{New} code (containing DC loops), we add two new flags. The first is {\tt -stdpar=gpu}, which enables DC loops to be parallelized and offloaded to the GPU\footnote{For {\tt nvfortran 21.7}, it appears that setting the {\tt -stdpar=gpu} flag implicitly sets the {\tt -acc=gpu} option as well.  This is an important consideration if one has {OpenACC} directives that should be ignored when using {\tt -stdpar}.}.  The other is {\tt -Minfo=stdpar} which outputs the compiler's parallelization messages (similar to {\tt -Minfo=accel}).  When using {\tt -stdpar=gpu}, unified managed memory is automatically enabled, making all allocatable arrays unified arrays.  This means the runtime is responsible for correct and efficient CPU-GPU data transfers during the run, and any OpenACC data movement directives on such arrays are essentially no-ops.  Static arrays are not made into unified arrays, so manual GPU data movement is still needed for good performance (note that {\tt diffuse} does not make use of any static arrays).  If one wants to continue to manage the GPU data manually (using OpenACC or OpenMP data movement directives), the option {\tt -gpu=nomanaged} can be used.

For the \emph{Experimental} code, since there are no directives, we simply use the standard parallelism option of {\tt -stdpar=gpu -gpu=ccXY,cudaX.Y}, and rely on the compiler to automatically detect the reductions and implement them correctly, as well as manage the GPU memory using unified memory.

For CPU parallelization, the \emph{Original} code has two implementations. One is to use OpenMP with the {\tt -mp} flag, and the other is to use OpenACC with the {\tt -acc=multicore} flag.  Even though the OpenMP compilation produces slightly better performance (as was shown in Sec.~\ref{sec:baseline}), we only use the OpenACC multi-core option.  This is because {\tt nvfortran} currently activates OpenACC when using {\tt -stdpar}, so we cannot use both {\tt -stdpar} for DC and OpenMP (as would be needed in the \emph{New} code) since OpenMP and OpenACC are not written to work together (and in the \emph{New} code case, causes a compiler error).  We note that when using OpenACC for multi-core CPU, the number of threads is controlled through the runtime variable {\tt ACC\_NUM\_CORES=<N>}, rather than OpenMP's {\tt OMP\_NUM\_THREADS=<N>}.
	 
For the \emph{Experimental} code, since there are no directives, we simply use the standard parallelism option of {\tt -stdpar=multicore}, relying on the compiler to automatically detect the reductions and implement them correctly.

\emph{\tt gfortran}: 
For GPU parallelization, the \emph{Original} code uses the flag {\tt -fopenacc}, which enables OpenACC directives.  In addition to this flag, the intended offload GPU must be specified. For NVIDIA GPUs, the flag {\tt -foffload=nvptx-none} is used (targeting specific compute capabilities is not currently implemented).  We also use the flag {\tt -fopenacc-dim=<DIM>} to specify the parallel topology for the offload kernels.  {\tt <DIM>} is set to three colon-separated values that map to ’gang’, ’worker’ and, ’vector’ sizes.   Since OpenACC supports acceleration for multiple GPU vendors, the default values for the topology may not be optimal.  Although this level of optimization is outside the scope of this paper, we observed that the {\tt nvfortran} compiler was using a vector length of 128 when compiling most OpenACC loops, so as a simple optimization, we use {\tt -fopenacc-dim=::128} for our tests.  The \emph{New} code is not supported on the GPU with {\tt gfortran} at this time.  This is because there is no current support for DC GPU offloading.

On the CPU, the \emph{Original} code uses {\tt -fopenmp}, which as above, activates OpenMP directives for multi-core CPU parallelism.  {\tt gfortran} does not support direct parallelism on DC loops.  Therefore, for the \emph{New} code, we must use {\tt gfortran}'s auto parallelization feature using the {\tt -ftree-parallelize-loops=<N>} flag, where {\tt <N>} is the number of threads to run on.  This auto parallelization analyzes both {\tt do} and DC loops and determines if they can be parallelized and if so, implements the parallelism.  Therefore, it can be used in the case of the \emph{New} code, as well as the \emph{Experimental} code.  Since the compiler is auto-analyzing the loops, it may detect the DC reduction loops and parallelize them correctly.

\emph{\tt ifort}: 
Since {\tt ifort} does not currently support GPU-offloading with DC or OpenACC, we only test it with CPU parallelism to ensure switching from directives to DC does not lose our CPU parallel capabilities when using {\tt ifort}.  For all code versions, we add the flags {\tt -fp-model precise} and {\tt -heap-arrays} as those are standard flags we use for runs of {\tt diffuse} to ensure robustness and precision, but they are not related to parallelization.   For the \emph{Original} code, we use the flag {\tt -fopenmp} in order to enable OpenMP directives to produce parallel code for multicore CPUs.   For the \emph{New} code, we use the same {\tt -fopenmp} flag as the \emph{Original} code, as it is also used to enable automatic parallelization of DC loops. The \emph{Experimental} code also uses the same {\tt -fopenmp} flag. However, as the documentation states that DC reduction loops are not supported, we do not expect {\tt ifort} to parallelize them, and rather run them in serial (although as will be shown, the current compiler version parallelizes the loops anyways, resulting in incorrect results).

\section{Results}
\label{sec:results}
Here we show timing results for all chosen compilers, code versions, and hardware (where supported).  Key questions we address are: 1) do the compilers that support GPU-acceleration with directives also support it using DC? 2) does replacing directives with DC lose CPU multicore parallelism? (i.e. do the compilers support DC for CPU multicore?) 3) for compiler-hardware combinations that support parallelizing DC, how does the performance compare to the baseline directive-based code?  We first report results for the \emph{New} code compared to the \emph{Original} code for each compiler and hardware type, and afterwards discuss results for the \emph{Experimental} code.

For each configuration, we run the test case of Sec.~\ref{sec:testcase}, and use the Linux program {\tt time} to record three times: {\tt real}, {\tt user}, and {\tt system}. The {\tt real} time is the wall clock time the code took to run. The {\tt user} time is the sum of all thread times, or how much total CPU computation time was spent. Using multiple threads should result in a lower {\tt real} time, but a (much) higher {\tt user} time. The {\tt system} time is the operating system overhead, which can include CPU-GPU data transfer, as well as other overheads.  All reported timings are averaged over 10 runs.

\subsection{Results using {\tt nvfortran}}
\label{sec:results_nv}
The results for the \emph{Original} and \emph{New} code run on the GPU with {\tt nvfortran} are shown in Table~\ref{section:Bridges2_NV_GPU_n}.  
\begin{table}[htbp]
	\vspace{-4mm}
	\begin{center}
		\caption{GPU timing results with {\tt nvfortran}.  Both runs used the additional compiler flag {\tt -gpu=cc80,cuda11.4}} 
		\label{section:Bridges2_NV_GPU_n}
		\begin{tabular}{||C{16mm} | C{38mm}|R{16mm} R{18mm} R{18mm}  ||} 
			\hline
			Code  & Compiler flags & real  (s)  & user   (s)  & system   (s)  \\ 
			\specialrule{.1em}{.05em}{.05em}
			\emph{Original} & {\tt -acc=gpu} & 35.07	& 34.46	& 0.59 \hspace*{1pt}\\
			\hline
			\emph{New} & {\tt -acc=gpu -stdpar=gp}  & 35.67	& 35.01 &	0.54 \hspace*{1pt}\\
			\hline
		\end{tabular}
	\end{center}
	\vspace{-8mm}
\end{table}
The time difference between the \emph{Original} code and the \emph{New} code is less than 2\%, and the standard deviation over the 10 runs is around ${\pm 0.1s}$ for both.  The slight increase in run time for the \emph{New} code is possibly due to its use of unified memory, which can be less efficient than manually managing the GPU-CPU memory as is done through OpenACC data directives in the \emph{Original} code.  This result does not achieve a full replacement of directives with DC since not all directives were replaced in the \emph{New} code.  However, the vast majority of them were, with only a few remaining directives on the reduction loops, showing great progress in replacing directives.

To ensure that we did not loose CPU multicore parallelism, we show the CPU results in Table~\ref{section:Bridges2_NV_CPU_n}.
\begin{table}[htbp]
	\vspace{-4mm}
	\begin{center}
		\caption{CPU timing results with {\tt nvfortran}} 
		\label{section:Bridges2_NV_CPU_n}
		\begin{tabular}{||C{16mm} | C{31mm}|R{16mm} R{18mm} R{18mm}  ||} 
			\hline
			Code  & Compiler flags  & real  (s)  & user   (s)  & system   (s)  \\ 
			\specialrule{.1em}{.05em}{.05em}
			\emph{Serial} & & 1284.59	&1272.45	&0.22 \hspace*{1pt}\\
			\hline
			\emph{Original} & {\tt -acc=multicore} & 224.18	&26214.96&	1965.06 \hspace*{1pt}\\
			\hline
			\multirow{2}{*}{\emph{New}} & {\tt -acc=multicore} &\multirow{2}{*}{219.57		}& \multirow{2}{*}{25638.37}&\multirow{2}{*}{1889.20 \hspace*{1pt}}\\
			&{\tt -stdpar=multicore}  &  &  & \\
			\hline
		\end{tabular}
	\end{center}
	\vspace{-8mm}
\end{table}
We see that, like in the GPU case, replacing directives with DC yields similar runtimes to the original code.  Here, the \emph{New} code with DC runs around 3\% faster than the \emph{Original} code, but both are within the standard deviation (${\pm 15s}$) of the 10 runs.  Therefore, there is no loss of CPU portability when using DC with {\tt nvfortran} for our mini-app.

\subsection{Results using {\tt gfortran}}
\label{sec:results_gf}
The result for the \emph{Original} code run on the GPU with {\tt gfortran} is shown in Table~\ref{section:Bridges2_NV_GPU_n}.
\begin{table}[htbp]
	\vspace{-4mm}
	\begin{center}
		\caption{GPU timing results with {\tt gfortran}.} 
		\label{ssection:Bridges2_GF_GPU_n}
		\begin{tabular}{||C{16mm} | C{48mm}|R{16mm} R{18mm} R{18mm}  ||} 
			\hline
			Code  & Compiler flags  & real  (s)  & user   (s)  & system   (s)  \\ 
			\specialrule{.1em}{.05em}{.05em}
			\multirow{2}{*}{\emph{Original}} & {\tt -fopenacc -foffload=nvptx-none}  & \multirow{2}{*}{49.52}	&\multirow{2}{*}{48.90}	&\multirow{2}{*}{0.54 \hspace*{1pt}}\\
			& {\tt -fopenacc-dim=::128} & & & \\
			\hline
			\emph{New} & No Support & - & - & - \hspace*{1pt}\\
			\hline
		\end{tabular}
	\end{center}
	\vspace{-8mm}
\end{table}
Unlike {\tt nvfortran}, {\tt gfortran} does not support GPU-acceleration using DC, nor is there auto parallelization support for GPU offloading.  Therefore, replacing the directives with DC currently breaks support for GPU-acceleration with {\tt gfortran}.  For NVIDIA GPUs, this is not a prohibitive limitation since the {\tt nvfortran} compiler is freely available.  However, for other accelerators (namely AMD GPUs), this loss of support may rule out using DC at this time.

Unlike for GPU-acceleration, CPU multi-core parallelism with {\tt gfortran} is not lost with DC, even though there is no direct support for DC parallelization. In Table \ref{section:Bridges2_GF_CPU_n}, we show the CPU timing results of the \emph{Original} and \emph{New} codes.  
\begin{table}[htbp]
	\vspace{-4mm}
	\begin{center}
		\caption{CPU timing results with {\tt gfortran}} 
		\label{section:Bridges2_GF_CPU_n}
		\begin{tabular}{||C{16mm} | C{48mm}|R{16mm} R{16mm} R{16mm}  ||} 
			\hline
			Code  & Compiler flags  & real  (s)  & user   (s)  & system (s)  \\ 
			\specialrule{.1em}{.05em}{.05em}
			\emph{Serial} & &1308.75 &1296.74 &0.16 \hspace*{1pt}\\
			\hline
			\emph{Original} & {\tt -fopenmp} & 191.90 &24117.72	&8.02 \hspace*{1pt}\\
			\hline
			\multirow{2}{*}{\emph{New}} & {\tt -fopenmp}  & \multirow{2}{*}{212.64} & \multirow{2}{*}{26588.59} & \multirow{2}{*}{8.65 \hspace*{1pt}}\\
			& {\tt -ftree-parallelize-loops=128}&&&\\
			\hline
		\end{tabular}
	\end{center}
	\vspace{-8mm}
\end{table}
We see that the performance difference between the codes is $\sim10\%$, within the standard deviations of the 10 runs (${\pm 13s}$ for the \emph{Original} and  ${\pm 18s}$ for the \emph{New} code).  The \emph{New} code is also able to be parallelized because  {\tt gfortran} treats DC loops as regular {\tt do} loops, which are parallelized using the auto parallelization feature.  However, the loops in {\tt diffuse} are fairly simple.  In other codes, the auto parallelization may not be able to handle more complex loops, that could otherwise be parallelized using directives.  Therefore, the result here should be viewed with caution.  Note also that the auto parallelization feature works the same on our code with regular {\tt do} loops as it does with DC loops (i.e. it would even parallelize the \emph{Serial} code version).

The thread control for the \emph{New} code is unique.  Since the OpenMP directives are still on the reduction loops, while the remaining loops use DC with no directives, the number of CPU threads used for the reductions is set by the standard {\tt OMP\_NUM\_THREADS} environment variable, while that used by the DC loops is controlled by the compiler flag value {\tt -ftree-parallelize-loops=<N>}.  This complicates the thread control, and also removes run-time thread control.

\subsection{Results using {\tt ifort}}
As mentioned in the introduction, the Intel OpenAPI Toolkit does not currently support GPU-acceleration using DC loops, but there are plans for support in the future. Therefore, here we focus on DC compatibility with multi-core CPU parallelism.  In Table \ref{section:Bridges2_IF_CPU_n}, we show the timing results for the \emph{Original} and \emph{New} codes.
\begin{table}[htbp]
	\vspace{-4mm}
	\begin{center}
		\caption{CPU timing results with {\tt ifort}.} 
		\label{section:Bridges2_IF_CPU_n}
		\begin{tabular}{||C{16mm} | C{22mm}|R{16mm} R{18mm} R{18mm}  ||} 
			\hline
			Code  & Compiler flags  & real  (s)  & user   (s)  & system   (s)  \\ 
			\specialrule{.1em}{.05em}{.05em}
			\emph{Serial} & & 1318.60 &1306.27 &0.18 \hspace*{1pt}\\
			\hline
			\emph{Original} & {\tt -fopenmp} & 194.86 &24213.11	& 320.53 \hspace*{1pt}\\
			\hline
			\emph{New} & {\tt -fopenmp} & 178.29	&21888.21 &280.65 \hspace*{1pt}\\
			\hline
		\end{tabular}
	\end{center}
	\vspace{-8mm}
\end{table}
We see that replacing directives with DC still allows for multi-core CPU parallelism, and surprisingly exhibits a nearly 10\% improvement in performance.  The standard deviation of the 10 runs was roughly ${\pm 8s}$, so this performance increase is significant.  It may be attributed to more efficient optimizations being available to the compiler when using DC compared to OpenMP directives. 

Since the implementation of DC parallelism is connected to OpenMP (as indicated by the use of the {\tt -fopenmp} flag), the number of threads remains controlled by the standard environment variable {\tt OMP\_NUM\_THREADS} (or optionally at compile time with {\tt -par-num-threads=<N>} which overrides {\tt OMP\_NUM\_THREADS}).

\subsection {Experimental Results}
\label{sec:results_exp}
As mentioned in Sec.~\ref{sec:codelist}, the current Fortran standard does not have a way to indicate to the compiler that a DC loop requires reduction or atomic operations.  However some compilers have implemented code analysis methods to automatically detect and implement such operations.  Therefore, the \emph{Experimental} code, which represents the ideal scenario of replacing all directives with DC loops, may work with some compilers.

Using {\tt nvfortran}, we found that the code parallelized and ran correctly on both the GPU and CPU. It appears {\tt nvfortran} detects the reductions and implements them correctly for our code.  The run times are shown in Table~\ref{section:Bridges2_NV_E}.
\begin{table}[htbp]
	\vspace{-4mm}
	\begin{center}
		\caption{GPU and CPU timing results for the \emph{Experimental} code with {\tt nvfortran}.} 
		\label{section:Bridges2_NV_E}
		\begin{tabular}{||C{20mm} | C{17mm} | C{32mm}|R{13mm} R{13mm} R{17mm}  ||} 
			\hline
			Code  & CPU/GPU & Compiler flag  & real  (s)  & user   (s)  & system   (s)  \\ 
			\specialrule{.1em}{.05em}{.05em}
			\hline
			\multirow{2}{*}{\emph{Experimental}} & \multirow{2}{*}{GPU}& {\tt -stdpar=gpu}  & \multirow{2}{*}{35.63}	&\multirow{2}{*}{35.01}	&\multirow{2}{*}{0.53 \hspace*{1pt}}\\
			& & {\tt -gpu=cc80,cuda11.4} & & & \\
			\hline
			\emph{Experimental} & CPU& {\tt -stdpar=multicore} & 219.21	&25654.35&	1906.40 \hspace*{1pt}\\
			\hline
		\end{tabular}
	\end{center}
	\vspace{-8mm}
\end{table}
They are nearly identical to those of the \emph{New} code shown in Tables~\ref{section:Bridges2_NV_GPU_n} and \ref{section:Bridges2_NV_CPU_n} which is expected since only a few small loops used directives for reductions in the \emph{New} code.  This means that for {\tt diffuse}, we can use DC to fully eliminate all directives and not lose any CPU or GPU performance with {\tt nvfortran}.  We reiterate that more complicated codes may not yet work with zero directives for a variety of reasons including not detecting complicated reduction or atomic operations, not being compatible with in-lined function calls, and, for GPU-acceleration, not supporting automatic memory management with static arrays.

For {\tt gfortran}, we only test the code on the CPU since there is no support for DC GPU-offloading.  In this case, the \emph{Experimental} code ran correctly, implying  the auto-parallelization done by the compiler was able to detect the reductions and parallelize them.  The run time is shown in Table~\ref{section:Bridges2_GF_CPU_E} and is roughly 10\% slower comparable to the run time of the \emph{New} code in Table~\ref{section:Bridges2_GF_CPU_n}.  However, the times are nearly within the standard deviation of the 10 runs (${\pm 18s}$).
\begin{table}[htbp]
	\vspace{-4mm}
	\begin{center}
		\caption{CPU timing results for the \emph{Experimental} code with {\tt gfortran}.} 
		\label{section:Bridges2_GF_CPU_E}
		\begin{tabular}{||C{20mm} | C{48mm}|R{14mm} R{14mm} R{15mm}  ||} 
			\hline
			Code  & Compiler flags & real (s)  & user (s)  & system (s)  \\ 
			\specialrule{.1em}{.05em}{.05em}
			\hline
			\emph{Experimental} & {\tt -ftree-parallelize-loops=128} & 236.28	&29565.01&	10.08 \hspace*{1pt}\\
			\hline
		\end{tabular}
	\end{center}
	\vspace{-8mm}
\end{table}

Using {\tt ifort}, the \emph{Experimental} code compiled and ran, but did not give the correct results.  This is because {\tt ifort} does not support implicit reductions of DC loops, yet parallelized the loop anyways when we used the {\tt -fopenmp} flag,  Therefore, the resulting inherent race conditions produced incorrect results.

\section{Discussion}
\label{sec:discussion}
In this paper, we have used a mini-app code to explore the current status of replacing {\tt do} loops using directives with {\tt do concurrent} (DC) loops for accelerated computing.  The original code used OpenACC for GPU-acceleration when compiled with {\tt gfortran} or {\tt nvfortran}, and OpenMP for multi-core CPU parallelism when compiled with {\tt gfortran}, {\tt nvfortran}, or {\tt ifort}.  We modified the code to replace the directives with DC and used a test case to explore the resulting compatibility, portability, and performance, all with the newest available versions of the compilers.  

{\bf Compatibility:} 
We found that only {\tt nvfortran} currently supports GPU acceleration with DC, and therefore replacing the directives removed GPU support when using {\tt gfortran}.  Since {\tt nvfortran} is freely available, this is not an insurmountable problem when running on NVIDIA GPUs.  However, {\tt gfortran} also has AMD (and possible future Intel) GPU support, making this an important consideration.  The {\tt ifort} compiler does not currently support GPU-acceleration with DC, but as Intel has indicated plans to add this support soon, switching from OpenACC directives to DC may increase compatibility (as {\tt ifort} only supports OpenMP GPU offload, not OpenACC).

We also found that the current Fortran specification for DC lacks features that are needed to guarantee correct parallelization of all of our mini-app's parallelizable loops (specifically, loops with reductions).  Indeed, when {\tt ifort} attempted to parallelize our reduction loops for the CPU, it resulted in incorrect results.  In contrast, the {\tt nvfortran} compiler has implicit reduction detection of DC loops, allowing us to replace all directives with DC.  The next release of the Fortran standard (202x) will include an explicit `reduce' clause on DC, which, when implemented, should alleviate this issue.

Another compatibility concern is that we currently use OpenACC directives to manually control GPU-CPU memory management, and removing these could cause extreme loss of performance.  In the case of {\tt nvfortran}, since it automatically activates its unified memory management feature when compiling with DC GPU-acceleration, this issue is avoided.  However, unified memory is limited to allocatable arrays, so static arrays may still require data management directives.  

Using cutting-edge language and compiler features have a risk of breaking backward compatibility with older compilers.  In this paper we used the most recent versions of the compilers we could for the best support, but on some systems this is not always available.  Container frameworks like Singularity (as used here) can help mitigate this issue, however the frameworks are also not always available on all systems, and can sometimes be complicated to use for large scale simulations. 

{\bf Portability:}
A key consideration in replacing directives with DC for GPU acceleration was to see if, by doing so, we still maintain CPU multi-core parallelism (that we originally used OpenMP directives to achieve).  We found that {\tt nvfortran} and {\tt ifort} compilers directly support DC for multi-core CPU parallelism, while {\tt ifort} requires directives on loops with reductions for correctness.  With {\tt gfortran}, while there is no direct support for DC parallelism, the loops can still be parallelized using {\tt gfortran}'s auto parallelization feature.  With this feature, even reduction DC loops are correctly recognized and parallelized.  Thus, all three compilers we use are able to keep multi-core CPU parallelism when replacing directives with DC (with {\tt ifort} still requiring some on reduction loops).

{\bf Performance:}
Replacing directives with DC allows much cleaner looking code and robustness due to being part of the standard language.  However, this is only worth while if it also results in acceptable performance.   Through our timings, we found that in both the GPU and CPU cases, the performance of the code after replacing directives with DC was comparable to that of the original directive-based code, with some configurations improving performance slightly, and in others, decreasing slightly.  For GPU runs with DC, {\tt nvfortran}'s unified memory was used, and the resulting performance was comparable to using manual OpenACC data directives.  However, more complicated codes may not be as compatible with unified memory and/or may lose some performance using it.  

{\bf Summary:} 
With {\tt nvfortran}, we were able to remove and replace all directives in our code with DC, and achieve efficient CPU and GPU parallelism.  However, this relied on specific features of {\tt nvfortran} including implicitly detecting reductions and the use of unified managed memory.  In order to maintain cross compiler compatibility, we can continue to use OpenACC/OpenMP directives for reductions and data movement until equivalent standard language features are written and widely supported.  Even with the remaining directives, using DC has a large benefit, as the number of directives is decreased dramatically.  

Can Fortran's {\tt do concurrent} replace directives for accelerated computing?  With {\tt nvfortran} and NVIDIA GPUs, for some codes (such as ours) the answer is yes, and with no (or minimal) loss of performance.  With upcoming language features and compiler implementations, more complicated codes may also eventually be parallelized without directives, and do so with support across multiple compiler and hardware vendors.

\section*{Appendix}
\label{sec:appendix}
{\bf Singularity containers:} 
In order to test the latest compilers and to simplify setup of our library dependencies, we utilized Singularity containers.  These containers allow one to run software in a containerized environment on any compatible system using only the container file.

{\bf Singularity container setup:}  
Singularity containers were straight forward to setup and use for our timings.  Two methods were used to create them. For {\tt nvfortran} and {\tt ifort}, a docker image of NVIDIA HPC SDK or Intel OneAPI HPC Toolkit was used to create a Singularity sandbox. For {\tt gfortran}, a similar sandbox with Ubuntu 21.04 was created and then {\tt gfortran} was installed with the {\tt apt-get} command. Once the sandboxes were created, the dependent libraries were installed. A sandbox is treated like a virtual machine, allowing us to to edit and install new software into the container (note this requires {\tt sudo} privileges).  Once all the needed software is installed, the sandbox is converted to a {\tt .sif} file which can be copied and run (without {\tt sudo} privileges) on any other compatible machine with Singularity installed, but it can no longer be edited.  However, the container is able to modify files outside itself, allowing us to compile and run the test cases.   For GPU-accelerated runs, a special flag is needed when running the container depending on the vendor of GPU.  For NVIDIA GPUs, the flag is {\tt -{}-nv}, while for AMD GPUs, the flag is {\tt -{}-rocm}.  For more details on Singularity containers, see Ref.~\cite{kurtzer_sochat_bauer}.

{\bf Singularity performance tests:} 
Using containers can sometimes cause performance overhead.  To ensure that using the Singularity containers does not cause significant overhead in our case, we ran two test cases on both a bare metal setup and with a container with the same compiler version (in this case {\tt gfortran} 10.2).  Table~\ref{Singularity_performance} shows timings of the test run using both the \emph{Original} and \emph{Serial} codes described in Sec.~\ref{sec:codelist}.
\begin{table}[htbp]
		\vspace{-4mm}
		\caption{Timing results on a Bridges2 CPU compute node using {\tt gfortran} 10.2 bare metal and form within a Singularity Container} 
		\label{Singularity_performance}
		\begin{center}
			\begin{tabular}{||C{18mm} C{23mm}| R{18mm} R{18mm} R{18mm}||} 
				\hline
				 Code & Run method & real  (s)  & user   (s)  & system   (s)  \\ [0.5ex] 
				\specialrule{.1em}{.05em}{.05em}
				\multirow{2}{*}{\emph{Serial}}
				&Bare Metal &  1306.10 & 1294.30 &  0.154 \hspace*{1pt}\\
				&Singularity & 1300.43  & 1287.50  & 0.168 \hspace*{1pt}\\
					\hline			
				\multirow{2}{*}{\emph{Original}}
				&Bare Metal &  164.87 & 20782.32 & 5.935 \hspace*{1pt}\\
				&Singularity & 165.27 & 20777.85 & 7.248 \hspace*{1pt}\\
				\hline			
			\end{tabular}
		\end{center}
		\vspace{-8mm}
\end{table}
We see that the runs using the Singularity container perform nearly identical to those run on bare metal, allowing us to confidently use the containers for the runs in the paper.

{\bf Reproducibility package:}  
The results in this paper can be reproduced using our reproducibility package hosted publicly at Ref.~\cite{mikic_zoran_2021_5253520} and on our website\footnote{\url{www.predsci.com/papers/dc}}.  The package contains three Singularity containers (for {\tt gfortran}, {\tt nvfortran}, and {\tt ifort}), as well as all code versions, compiler options, and test cases.   The package requires minimal customization (only specifying hardware-specific compiler options) of the main script, which can then be used to automatically run either all, or a subset, of runs from the paper.  See the documentation in the package for more details.  A reference solution is also provided for validation.   Note that runs using GPU-acceleration require having an NVIDIA GPU with compatible drivers installed on the system.

\bibliographystyle{splncs04}
\bibliography{ref.bib}

\end{document}